# Along-trajectory acoustic signal variations observed during the hypersonic reentry of the OSIRIS-REx Sample Return Capsule

Elizabeth A. Silber[1,*], Daniel C. Bowman[1,2,**]

[1]Sandia National Laboratories, Albuquerque, NM, 87144, USA

[2]Pacific Northwest National Laboratory, Richland, WA, 99354, USA

[*]Corresponding author

E-mail: esilbe[at]sandia.gov

[**]Was at SNL when this work was completed

**Abstract**

The reentry of the OSIRIS-REx Sample Return Capsule (SRC) on September 24, 2023, presented a rare opportunity to study atmospheric entry dynamics through a dense network of ground-based infrasound sensors. As the first interplanetary capsule to reenter over the United States since Stardust in 2006, this event allowed for unprecedented observations of infrasound signals generated during hypersonic descent. We deployed 39 single-sensor stations across Nevada and Utah, strategically distributed to capture signals from distinct trajectory points. Infrasound data were analyzed to examine how signal amplitude and period vary with altitude and propagation path for a non-ablating hypersonic object with well-defined physical and aerodynamic properties. Raytracing simulations incorporated atmospheric specifications from the Ground-2-Space model to estimate source altitudes for observed signals. Results confirmed ballistic arrivals at all stations, with source altitudes ranging from 44 km to 62 km along the trajectory. Signal period and amplitude exhibited strong dependence on source altitude, with higher altitudes corresponding to lower amplitudes, longer periods, and reduced high-frequency content. Regression analysis demonstrated strong correlations between signal characteristics and both altitude and propagation geometry. Our results suggest, when attenuation is considered, the amplitude is primarily determined by the source, with the propagation path playing a secondary role over the distances examined. These findings emphasize the utility of controlled SRC reentries for advancing our understanding of natural meteoroid dynamics, refining atmospheric entry models, and improving methodologies for planetary defense. The OSIRIS-REx SRC campaign represents the most comprehensive infrasound study of a hypersonic reentry to date, showcasing the potential of coordinated geophysical observational networks for high-energy atmospheric phenomena, including space debris reentries.

**Keywords:** reentry, infrasound, infrasound propagation, meteoroids, remote sensing, shock wave, OSIRIS-REx, planetary defense

## Introduction

Earth's atmosphere encounters extraterrestrial material ranging from microscopic particles to meter-scale objects, with entry velocities spanning 11.2 to over 72 km/s (Ceplecha, et al. 1998). While smaller particles typically ablate entirely at high altitudes, larger meteoroids and asteroids deposit substantial energy along their ablational trajectories, generating shock waves (Bronshten 1983, Silber, et al. 2018). Meter-sized meteoroids represent an important middle ground, bridging the gap between ablation-dominated particles and larger impactors capable of reaching the surface. These objects present both a scientific and societal interest: they contribute to our understanding of planetary formation, atmospheric entry physics, and planetary defense strategies. However, our ability to perform detailed studies is hindered by the fact that these objects are statistically more rare than smaller, more numerous impactors. For instance, prediction windows are frequently limited to just a few hours (Jenniskens, et al. 2021), restricting the ability to plan coordinated observational campaigns ahead of time. Ground-based observational networks (Devillepoix, et al. 2020) are further constrained to landmasses, leaving large regions of the Earth unmonitored, and many detections rely on sensors designed for other purposes, such as U.S. government systems that observe meteoroids incidentally as secondary phenomena (Nemtchinov, et al. 1997, Silber 2024b).

Sample Return Capsules (SRCs) provide an unparalleled opportunity to complement studies of meteoroid events, offering predictable and well-characterized conditions that mitigate the challenges posed by the unpredictable nature and limited observational coverage of natural entries. These compact capsules, engineered for interplanetary missions, reenter Earth's atmosphere at velocities exceeding 11 km/s, comparable to the lower range of meteoroid speeds. Unlike meteoroids, which are highly stochastic in their trajectories, compositions, and entry angles, SRCs have well-known source parameters, follow predictable paths, experience negligible ablation, and do not fragment under nominal conditions (Silber, et al. 2023). Their reentry timing is known far in advance, which provides ample time for observational campaign preparation. Therefore, it is immensely beneficial to leverage controlled reentries with the aim to investigate shock wave formation, energy deposition, and infrasound propagation.

Since the Apollo program, there have been only five interplanetary SRC reentries recorded by geophysical instruments, with the OSIRIS-REx (Origins, Spectral Interpretation, Resource Identification, and Security–Regolith Explorer) SRC being the most recent (Carr, et al. 2024, Fernando, et al. 2024a, Fernando, et al. 2024b, Silber, et al. 2023, Silber, et al. 2024b). The prior reentries, Genesis (ReVelle, et al. 2005), Stardust (ReVelle, et al. 2006), and Hayabusa 1 and 2 (Ishihara, et al. 2012, Sansom, et al. 2022), have demonstrated that SRC reentries can be effectively captured by geophysical sensing techniques, particularly infrasound and seismic instrumentation. Infrasound, or a low frequency acoustic wave (< 20 Hz), is generated as a byproduct of a shock wave decay. It can propagate efficiently over long distances with minimal attenuation (Evans, et al. 1972), making it ideal for studying high-energy atmospheric events such as bolides (ReVelle 1976), rocket launches (Pilger, et al. 2021), controlled reentries (ReVelle, et al. 2005, Silber, et al. 2024b), and explosions (Bowman, et al. 2025).

The OSIRIS-REx SRC reentry, the first interplanetary capsule to return over the United States since Stardust in 2006, presented an exceptional opportunity for a comprehensive geophysical observational campaign. Launched by the National Aeronautics and Space Administration (NASA) on September 8, 2016, OSIRIS-REx was designed to collect and return samples from the near-Earth asteroid Bennu, a primitive body that contains clues about the early solar system (Lauretta, et al. 2017). After successfully collecting samples in 2020, the spacecraft approached Earth on September 24, 2023, releasing its SRC for reentry at 10:41:55 UTC. The capsule entered Earth's atmosphere at a velocity of 12.36 km/s and landed at the Utah Test and Training Range (UTTR) at 14:52:09 UTC (Francis, et al. 2024).

A distributed sensor network can capture acoustic signals from distinct trajectory points during hypersonic reentry (e.g., Silber 2024a). In particular, for hypersonic sources, the geometry of the Mach cone plays a critical role in signal propagation (Anderson 2000). For example, at Mach 30 (Mach number is the ratio of the object's speed and the local speed of sound), the Mach cone half-angle is only 1.9°, resulting in predominantly ballistic shocks and focused energy emission pattern (Anderson 2000, ReVelle 1976, Silber, et al. 2018). In near field (< 300 km), ballistic shocks typically travel the shortest path and arrive nearly perpendicular to the trajectory, making them dominant contributors to observed signals. Therefore, an infrasound sensor network positioned in the proximity of the direct path of a high-altitude moving source could, in principle, capture signals generated at distinct points along the trajectory, as for the first time practically demonstrated through observations of an earthgrazing fireball (Silber, et al. 2024a). Building on this theoretical framework, we deployed a dense sensor network over a wide geographical region to experimentally investigate the extent to which signals can be captured from distinct points along the trajectory and how signal properties, such as period and amplitude, vary with altitude and propagation geometry. Our study demonstrates the potential of controlled SRC reentries to advance our understanding of natural meteoroid dynamics, refine shock wave propagation models, and improve methods for planetary defense and atmospheric monitoring.

Here, we present the analysis of infrasound signatures from the OSIRIS-REx SRC reentry, recorded by a dense network of ground-based sensors spanning Nevada and Utah, USA, as part of the largest geophysical campaign for a hypersonic reentry to date (Carr, et al. 2024, Fernando, et al. 2024a, Fernando, et al. 2024b, Silber, et al. 2023, Silber, et al. 2024b). By leveraging the predictable dynamics of a controlled reentry, this study provides a unique framework for examining how infrasonic signal parameters, such as amplitude and period, depend on source altitude, SRC trajectory geometry, and propagation path.

**Methods**

A comprehensive geophysical observational campaign was prepared in advance of the OSIRIS-REx SRC reentry (Silber, et al. 2024b). Leveraging NASA's Entry, Descent, and Landing (EDL) trajectory data (Francis, et al. 2024), sensors were strategically deployed to maximize the detection of geophysical signals. Instrument placement focused on two key regions: the vicinity of Eureka, NV, corresponding to the predicted location of peak heating, and the Nevada-Utah state line, approximately 200 km downrange (Figure 1). This latter region included the SRC's transition from hypersonic to supersonic flight and the subsequent subsonic "dark flight" phase before landing at the Utah Test and Training Range (UTTR) (Ajluni, et al. 2015, Francis, et al. 2024, Silber, et al. 2024b). Deployment details are thoroughly discussed in Silber, et al. (2024b); here, we summarize the most relevant aspects.

The analysis focuses on infrasound signals recorded by a ground-based network of 39 single-sensor stations. All stations were equipped with the same sensors (Gem infrasound microbarometers) to eliminate potential instrument bias. These portable, low-power sensors operate at a sampling rate of 100 Hz (Anderson, et al. 2018). The network was arranged into three distinct lines, and was designed to target specific regions of interest while remaining logistically feasible (Figure 1, Table S1):

- T Line: Positioned nearly beneath the trajectory to capture near-vertical arrivals from the hypersonic flight path.

- A Line: An orthogonal transect near Eureka, NV, positioned to capture signals associated with the region of maximum energy deposition during peak heating.

- C Line: A second orthogonal transect near the Nevada-Utah state line, designed to observe signal propagation up to dark flight (i.e., the SRC becomes subsonic and is no longer luminous).

The SRC entered the Earth's atmosphere at 14:41:55.57 UTC above San Francisco, CA at a shallow angle of 8.2° (±0.08°) and a velocity of 12.36 km/s (Ajluni, et al. 2015, Francis, et al. 2024). The atmospheric interface was at 125 km over the equatorial radius, which corresponds to a geodesic altitude of ~132.89 km. The SRC followed a southwest-to-northeast trajectory and landed at the UTTR. Due to a lack of detailed telemetry during the SRC descent, real-time trajectory information is unavailable. NASA is currently in the process of reconstructing the trajectory based on limited observational data. At present, a modeled post-flight reconstruction of the Entry, Descent, and Landing (EDL) trajectory has been made available (Francis, et al. 2024).

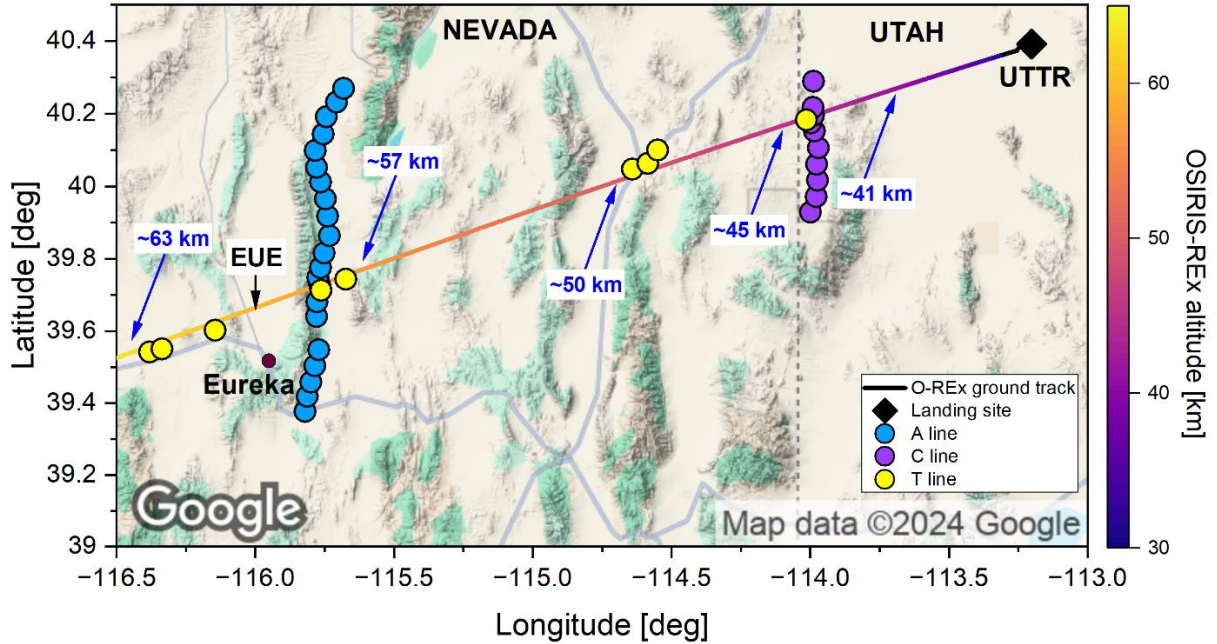

Figure 1: Map showing the deployment regions of ground-based single infrasound sensor stations. The Eureka Airport (EUE) is shown with an arrow. The SRC travelled from southwest to northeast. The labels with blue arrows show the approximate SRC altitude at various points along the trajectory. Full color figure is available in an online version of the article.

The first step was to identify infrasound detections on all stations, which proved to be relatively straightforward since the signals were readily apparent in raw waveforms. The recorded time-series data were processed using a Butterworth filter to isolate the infrasound signals from background noise. Details about this approach for processing and analyzing signals from high-altitude hypervelocity sources can be found in Silber (2024a). Given the broadband nature of the acoustic energy, data were filtered using a fourth-order, zero-phase Butterworth filter across multiple frequency bands to analyze signal characteristics comprehensively (Figures S1-S3). The lower frequency cutoffs were varied between 0.1 Hz and 1 Hz (0.1, 0.2, 0.25, 0.3, 0.4, 0.5, 1 Hz), and the upper frequency cutoffs were varied between 20 Hz and 45 Hz (20, 25, 30, 40, 45 Hz). For each frequency band, we measured the following signal parameters: maximum amplitude, peak-to-peak amplitude, and signal period. Analyzing signals across multiple frequency bands enabled us to assess the sensitivity of period and amplitude measurements to filter bounds, and to quantify uncertainties in the derived signal parameters.

Due to the short signal duration, along with shock wave attenuation and relaxation influencing negative phase features, we developed a robust and repeatable method for signal period measurement tailored to the characteristics of the recorded signals. Specifically, the period was determined using zero crossings at the point of maximum amplitude, focusing on the first half-cycle, as it consistently exhibited the most well-defined and stable signal structure. This value was then multiplied by two to estimate the full-cycle period. The same methodology was uniformly applied across all recorded signals and frequency bands, ensuring consistency, comparability, and reproducibility in the measurements.

Propagation modeling was conducted using InfraGA, an open source computational ray-tracing tool for simulating infrasound propagation in a three-dimensional, spherical Earth framework (Blom 2014). Using geometric ray-tracing techniques, it calculates acoustic paths, including eigenrays that connect the source to the receiver (Blom, et al. 2017). InfraGA incorporates realistic atmospheric parameters (e.g., temperature, pressure, wind), enabling detailed simulations over regional and global distances and accounting for the Earth's curvature. We used the Ground-2-Space (G2S) atmospheric specifications (Hetzer 2024) corresponding to the hour of the OSIRIS-REx SRC reentry, extracted for two locations: Eureka, NV (39.7480°N, -115.7730°E), near the peak heating point, and the Nevada-Utah state line, further downrange (40.1052°N, -113.9700°E) (Figure S4). G2S combines real-time observational data and climatological models to generate layered profiles of temperature, pressure, and wind from the surface to the lower thermosphere (Drob, et al. 2003).

We used the EDL trajectory of the SRC with propagation modeling to systematically search for all potential eigenrays connecting the source to each infrasound station. This was inspired by the approach previously used in the context of meteor infrasound (Brown, et al. 2011, Silber 2024a, Silber, et al. 2014). The eigenray search parameter space extended from -10° (nearly horizontal) to -89° (nearly vertically downward). Rays were traced from approximately 2000 densely spaced points along a linear segment of the trajectory. Each point was treated as an independent source. This accounts for variations in source altitude, trajectory geometry, and atmospheric conditions, and enables a comprehensive representation of potential signal arrivals at each station.

To estimate the most probable source altitudes, the modeled propagation paths and signal travel times were compared with the observed signal travel times and station distances from the SRC trajectory segment (3D distance and ground track distance). Since a hypersonic source on a linear trajectory is likely to produce ballistic shocks, which typically arrive at a near-perpendicular angle to the trajectory and follow the shortest path, we collected all eigenrays predicted to arrive at such geometry. It should be noted that inherent uncertainties might persist due to atmospheric variability, as G2S profiles may not fully resolve transient or localized conditions during the event. Additionally, minor inaccuracies in the EDL trajectory could introduce errors in source location inputs. Cross-validation against observations from multiple stations mitigates these effects and improves the reliability of the results.

Figure 2 illustrates the method used to associate eigenrays with a potential shock source, using station A19 as an example. All stations were processed using the same approach. The three panels show the relationships between the predicted signal travel time and the parameters of interest for all existing eigenrays: altitude (Figure 2a), ground distance (Figure 2b), and total 3D distance (Figure 2c). Although the results include a larger portion of the trajectory (travel times >245 s), the plots are zoomed in to the relevant region for clarity. The vertex in each panel, denoted by the red star, corresponds to the ballistic arrival, which represents the shortest possible travel time for the shock wave to reach the station. In Figure 2a, the source altitude is inferred as 56.2 km. The symmetry in the ray paths around the vertex indicates the consistency of raytracing with ballistic arrivals. Yellow points indicate a one-second uncertainty in travel time, which translates to an altitude uncertainty of approximately ±0.75 km. In Figure 2b, the altitude is consistent with that inferred in Figure 2a (56.4 km), emphasizing agreement between ground distance and travel time results. In Figure 2c, similar to Figures 2a-b, the results reaffirm the inferred source altitude of approximately 56.1 km. In the bottom two panels, the color gradient, ranging from lower altitudes (green) to higher altitudes (orange), reflects the altitudinal variation of eigenrays between 51 and 60 km altitude (the altitude range probed with raytracing extended 50 – 75 km for this station). To validate the consistency of this approach, the predicted arrival times were compared with the observed arrival times across all stations, showing excellent agreement. Finally, to derive the most probable source altitude and the associated error, we collated all eigenrays arriving within the uncertainty window (see Figure 2a) and calculated mean and standard deviation.

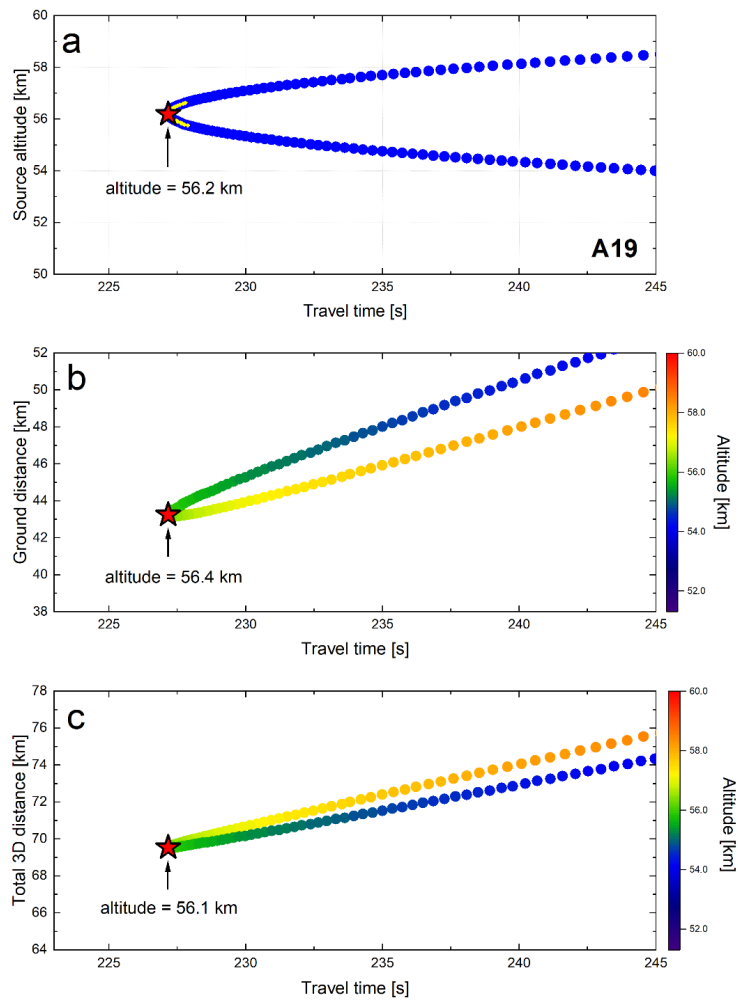

Figure 2: Raytracing results for station A19. Each dot represents an eigenray solution from a discrete point along the SRC trajectory, with the abscissa indicating the predicted travel time for all possible arrivals. The red star marks the eigenray associated with the ballistic arrival. The ordinate is the (a) source altitude, (b) source-to-station ground distance, and (c) source-to-station 3D distance. Full color figure is available in an online version of the article.

InfraGA outputs attenuation in the form of geometric spreading ($T_g$) and absorption ($T_a$) in decibels (dB) for each eigenray. To account for propagation effects, we used these attenuation factors to convert the signals observed on the ground to their effective amplitude ($A_s$) at 1 km from the source. We conducted four suites of propagation modeling simulations at central frequencies of 1, 5, 10, and 20 Hz. The frequencies of 5 Hz and 10 Hz encompass the infrasonic range of the SRC's dominant frequencies. The correction factor ($A_{cf}$) used to convert total attenuation ($T_{att}$) from dB to linear amplitude is given by $A_{cf} = 10^{(T_{att}/20)}$, where $T_{att}$ is the combined effect of geometric spreading and atmospheric absorption, i.e., $T_{att} = T_g + T_a$. The source amplitude is then calculated as $A_s = A_{obs}/A_{cf}$, with units of pascals. To account only for geometric spreading, the conversion becomes $A_s = A_{obs}/10^{(T_g/20)}$. We note that the gain in pressure amplitude due to the specific acoustic impedance contrast as the signal travels from the SRC altitude to the ground nearly cancels out geometric attenuation in these cases.

## Results

The recorded signals exhibit a characteristic N-wave signature (Dumond, et al. 1946), with a prominent initial half-cycle followed by an extended and more dispersed negative phase. Additionally, a coda is present following the main signal, possibly arising from atmospheric scattering or reflections. An apparent precursor energy seen in the signals is a digitizer artifact. Figure 3 shows representative filtered signals recorded on sensors along the T line, which was situated nearly vertically beneath the SRC's flight path. The lateral offset of the stations from the trajectory ranged between 1.1 and 2.2 km (Table S1). Relative ground distances (west to east) from T1 to other stations are as follows: 56.5 km (T7), 64.7 km (T8), 159.4 km (T10), 164.4 km (T11), and 215.9 (C7). Notably, the earliest signal was recorded at C7 and the latest at T1, opposite the flight direction. This reversal is attributed to the SRC's lowest altitude during the overflight near C7 (~44 km) and highest at T1 (~62 km), resulting in a shorter propagation distance for the signals to reach the eastern stations. Notably, the signal amplitude is evidently lower and the period longer for higher altitudes, as observed near T1, while signals originating from lower altitudes, such as near C7, exhibit higher amplitude and shorter periods, and a greater contribution of higher frequencies. The signal measurements for all stations are given in Table S2 (supplemental materials). The signal properties for all the stations are listed as measured.

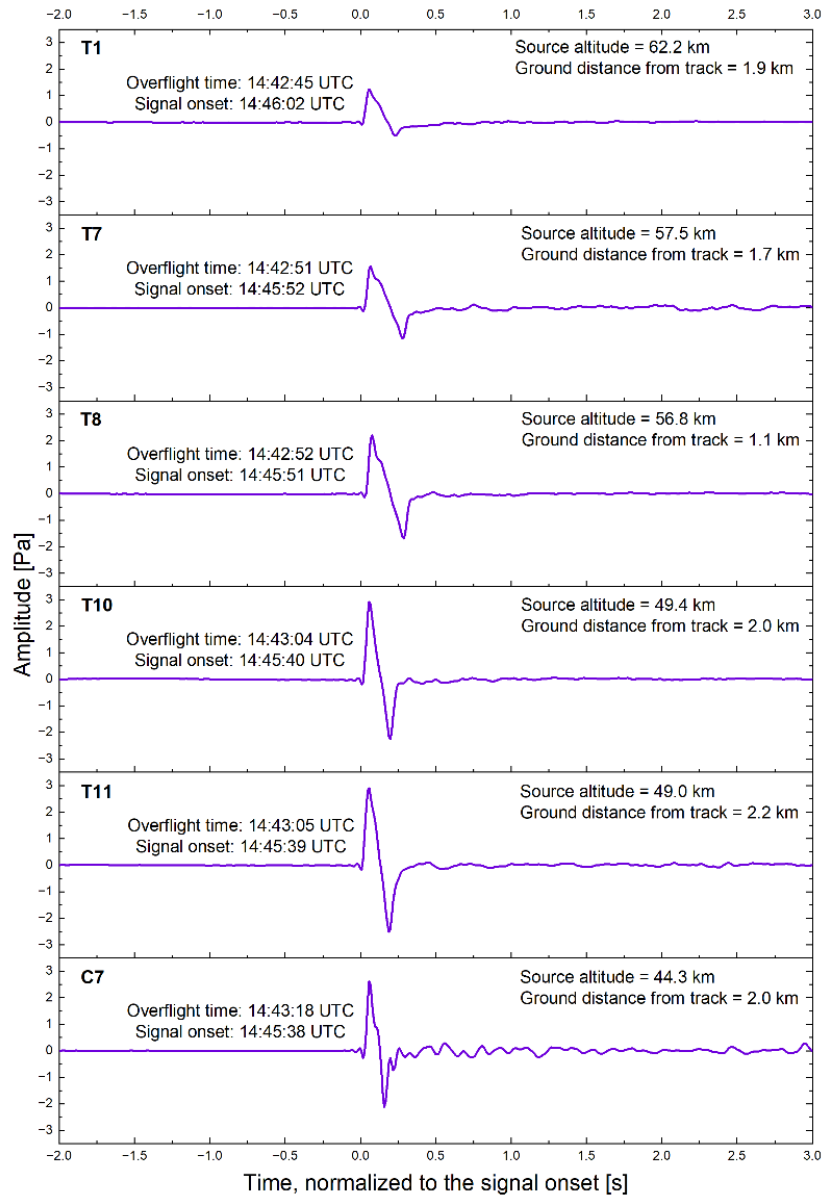

Figure 3: Infrasound signals received on stations beneath the trajectory, from westmost point (top panel) to the eastmost point (bottom panel). For clarity, the time axis is scaled such that the signal onset time is zero seconds. The actual arrival times and SRC overflight times are annotated on each panel. The signals shown here were processed using a Butterworth filter with cutoff frequencies of 0.25 to 25 Hz. The stations are arranged sequentially from the westernmost station (T1) to the easternmost station (C7), with a total ground separation of 215.9 km between them. An apparent precursor 'ripple' seen in the signals is due to the digitizer effect in Gems.

Raytracing results confirm that the observed infrasound arrivals across all stations align remarkably well with ballistic shock predictions, having virtually one-to-one (to a fraction of a second) correspondence to the EDL trajectory timing at derived altitudes. This alignment persists throughout the entire trajectory segment sampled by the infrasound network, spanning altitudes from 62.2 km down to 43.7 km. This agreement reflects the self-consistency of the raytracing methodology and supports the hypothesis that the recorded signals are dominated by ballistic arrivals from hypersonic shock sources. This enabled a reconstruction of the overflight times associated with specific altitudes, further reinforcing the reliability of the inferred shock source altitudes. The source altitudes derived through raytracing are given in Table S3 (supplemental materials).

Figure 4 provides a 3D visualization of eigenrays for two representative stations from each of the three sensor lines: Line A (Figures 4a and 4b), Line C (Figures 4c and 4d), and Line T (Figures 4e and 4f). Eigenrays are color-coded according to travel time, with yellow representing the shortest times corresponding to ballistic arrivals and dark blue indicating longer travel times (and consequently non-ballistic arrivals). In each panel, the most probable source altitude, which is also associated with the ballistic shock arrival, is denoted with a red star. For clarity, only every 10th eigenray is displayed. We also supply gif animations with 3D results in the supplemental materials. The gradient from faster (yellow) to slower (blue) travel times reflects altitudinal and temporal variations in eigenray paths, with the color scale adjusted for each station to capture trajectory-specific differences.

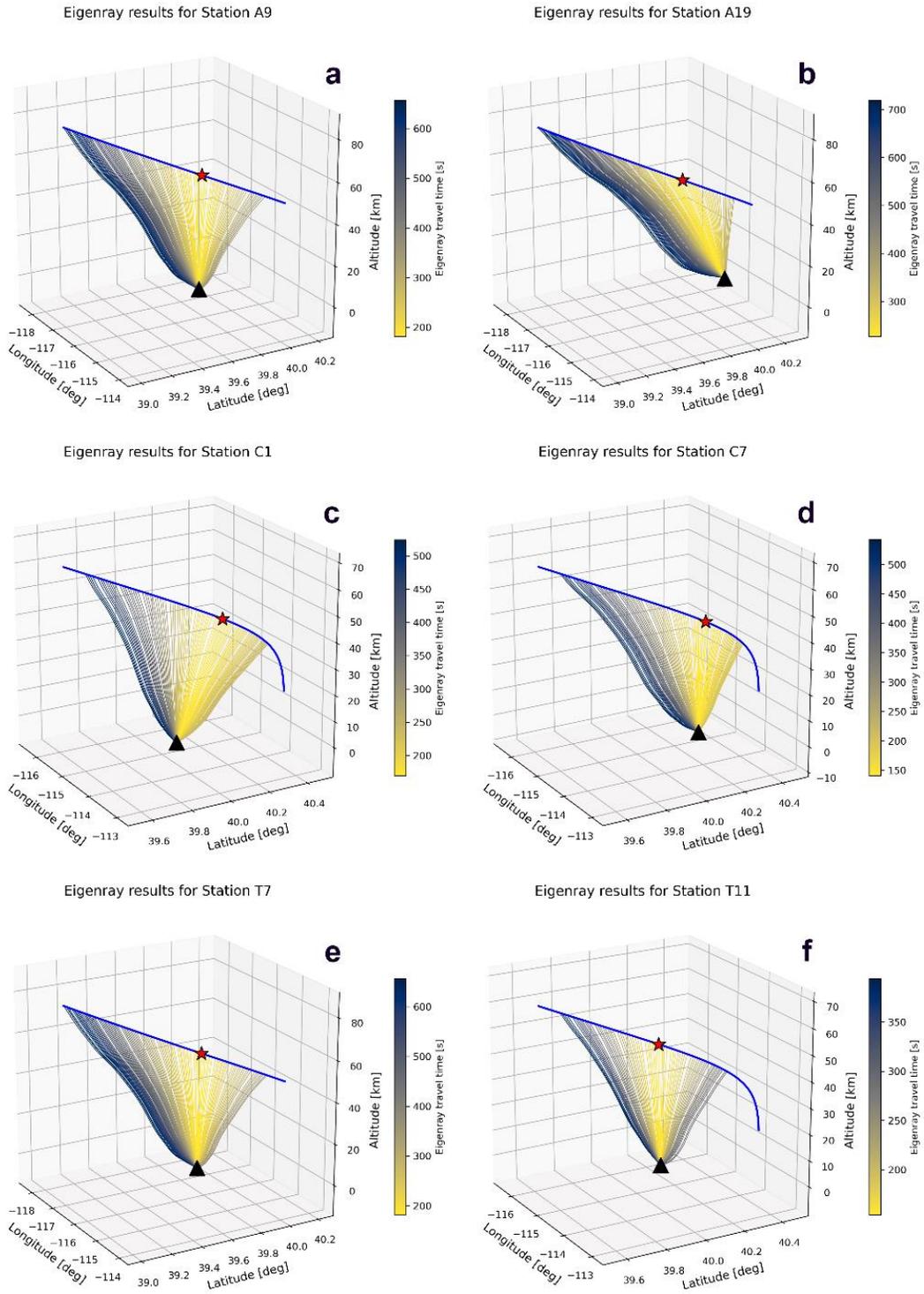

Figure 4: Raytracing results for two representative stations from each of the three lines (A, C, and T). Only every 10th ray is plotted for clarity. Note that the colorbar scale varies from panel to panel. The most probable source of the shock is denoted with a red star. Full color figure is available in an online version of the article.

Having derived the most probable source altitudes, we plotted these points as a function of sensor distance along the SRC ground track (Figure 5a) and station arrangement across three sensor lines (A, T, and C) (Figure 5b). The data points are color-coded according to the signal peak-to-peak amplitude, indicating the relationship between altitude, geographic distribution, and signal amplitude. Figure 5 demonstrates that each sensor captured signals from a slightly different part of the SRC's trail. The steady decrease in source altitude with increasing ground track distance reflects the gradual descent of the SRC along its trajectory (Figure 5a). Sensors closer to Eureka, NV captured signals originating from higher-altitude portions of the trail (~62 to 55 km), while those near the Nevada-Utah border recorded signals from lower-altitude regions (~50 to 44 km). The amplitude variation, with lower-altitude signals exhibiting higher peak-to-peak amplitudes, further emphasizes the distinct phases of the SRC's flight path captured by the network. The arrangement of stations across the three sensor lines (A, T, and C) illustrates the spatial distribution of captured signals (Figure 5b). Stations in the Nevada region recorded signals corresponding to the earlier, higher-altitude segments of the trail, while stations near the Nevada-Utah border captured signals from later, lower-altitude segments. The color-coded amplitudes show a consistent pattern: signals from higher altitudes have lower amplitudes, while those from lower altitudes have higher amplitudes.

Figure 6 shows the relationship between signal peak-to-peak amplitude and signal period, with data points color-coded according to the inferred source altitude. The altitude data provides the context into how the shock source height affects signal characteristics. The signals cluster into two distinct groups corresponding to the two primary deployment regions: (1) high-altitude sources near the peak heating region (~61 km), and (2) lower-altitude sources near the maximum dynamic pressure region (~44 km). Signals from higher altitudes (Eureka, NV) exhibit longer periods and lower amplitudes, forming a cluster in the lower-right portion of the plot. In contrast, signals originating from lower altitudes (UT-NV) have shorter periods and higher amplitudes, forming a cluster in the upper-left region.

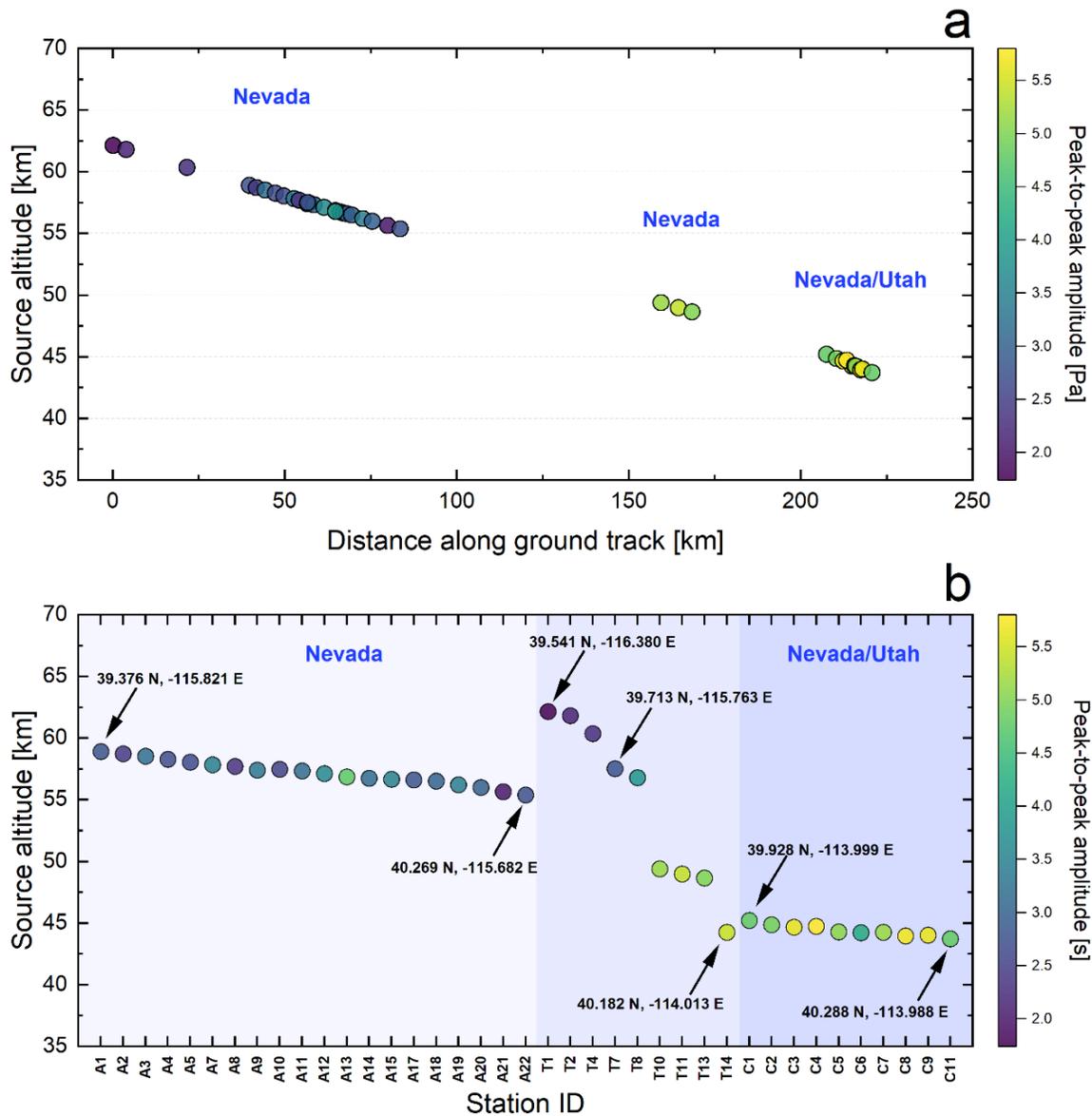

Figure 5: Source altitude as derived through propagation modeling plotted against (a) sensor distance along the SRC ground track and (b) stations (Lines A, T, and C). The origin in panel (a) corresponds to Station T1. The geographic coordinates for representative stations are annotated in panel (b). All data points are color coded according to the signal peak-to-peak amplitude in pascals. Full color figure is available in an online version of the article.

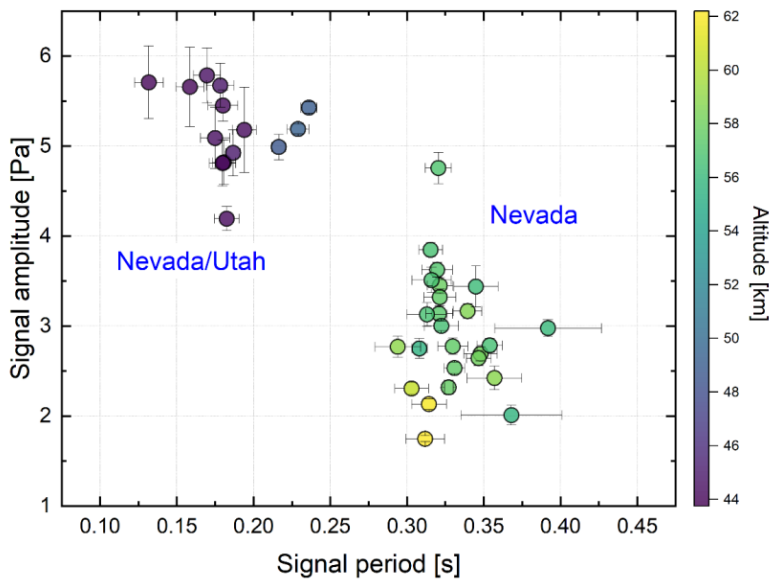

Figure 6: Relationship between signal peak-to-peak amplitude and signal period, with data points color-coded by derived source altitude. Signals from higher altitudes exhibit longer periods and lower amplitudes, while signals from lower altitudes are characterized by shorter periods and higher amplitudes. Error bars represent uncertainties in period and amplitude measurements (see Table S2). Full color figure is available in an online version of the article.

We performed regression analysis using the software package OriginPro® to evaluate the sensitivity of signal period and amplitude to source altitude and distance from the trajectory (Figure 7, Table S4). We investigated the relationships between signal period and amplitude and the following parameters: source altitude (Figure 7a and 7b), total 3D distance from source to sensor (Figure 7c and 7d), and ground distance from the trajectory to sensor (Figure 7e and 7f). The equations and regression statistics, including Pearson's $r$, $R^2$ coefficient of determination (COD) and adjusted $R^2$, are provided within each panel and in Table S4 (supplemental materials). The regression analyses show that source altitude and total 3D distance are the dominant factors influencing signal amplitude and period, with high correlations in Figure 7a through Figure 7d. A clear positive correlation is observed between signal period and source altitude ($R^2$ = 0.85, Pearson's $r$ = 0.92) and a negative correlation between amplitude and source altitude ($R^2$ = 0.80, Pearson's $r$ = -0.89). In terms of the total 3D distance, there is a positive correlation between the signal period and 3D distance ($R^2$ = 0.73, Pearson's $r$ = 0.85), and a negative, albeit weaker, correlation between amplitude and SAND2025-03153O3D distance ($R^2$ = 0.64, Pearson's $r$ = -0.80). Finally, there is a weak correlation between signal period ($R^2$ = 0.17, Pearson's $r$ = 0.41) and amplitude ($R^2$ = 0.11, Pearson's $r$ = -0.34), respectively, and ground distance from the trajectory. The source amplitudes corrected for attenuation show similar trends.

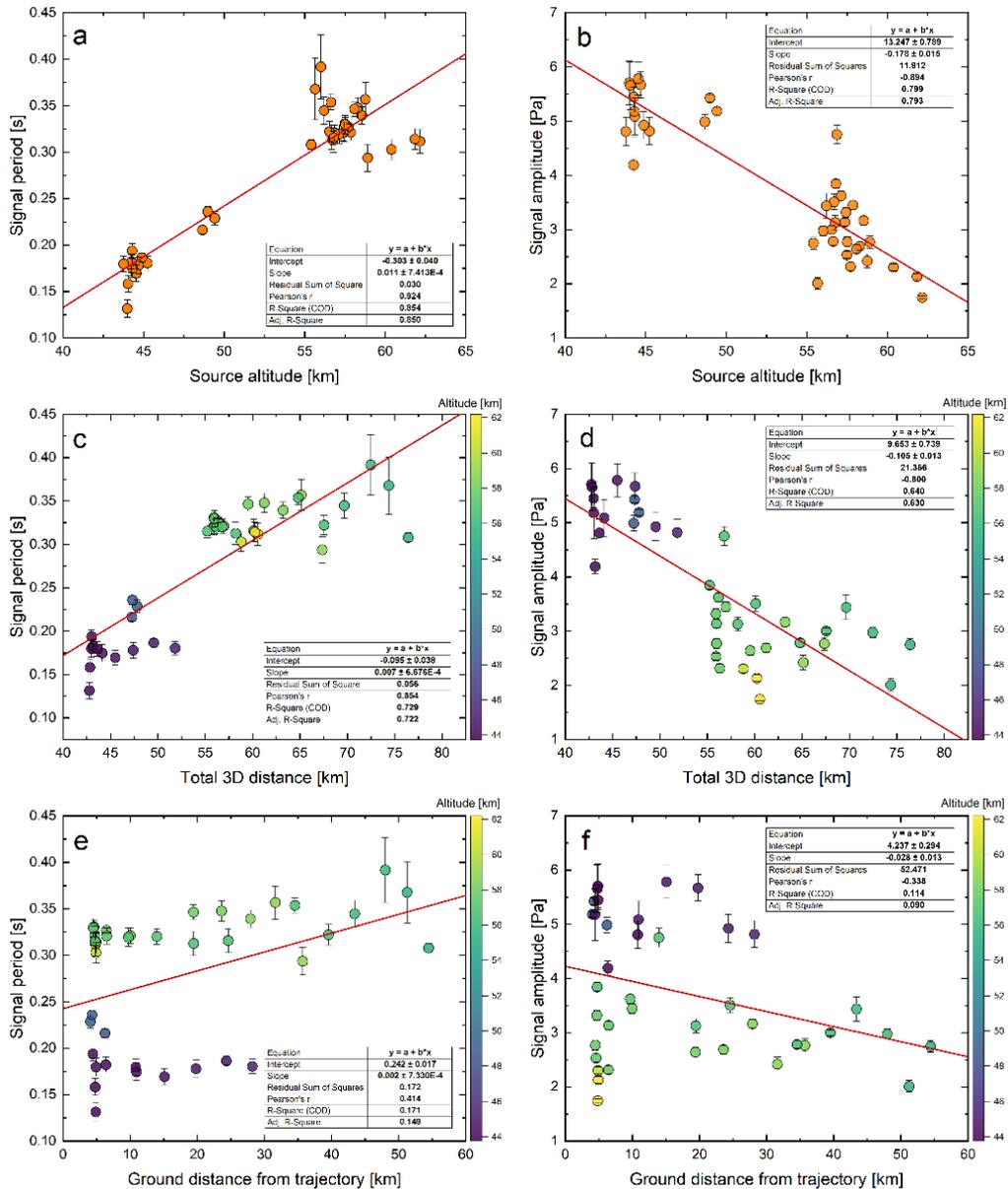

Figure 7: Signal period and amplitude as functions of source altitude (a,b), total 3D distance (c,d), and ground distance from the trajectory (e,f), with linear regression analyses shown in all panels. Data points are color-coded according to source altitude, with error bars representing uncertainties in signal measurements. Regression statistics, including slope, intercept, Pearson's $r$ and $R^2$, are provided within each panel and also listed in Table S5. Full color figure is available in an online version of the article.

## Discussion

Before we delve into the discussion and implication of our results, it is worth mentioning the fundamental differences between shock waves produced by ablating bodies (e.g., meteoroids) and non-ablating bodies (e.g., SRCs and similar artificial objects). Unlike SRCs, meteoroids, composed of naturally occurring materials with varying porosity, composition, and structural integrity, undergo intense ablation and often fragmentation during atmospheric entry (Bronshten 1983, Ceplecha, et al. 1998, Romig 1964). Ablation-driven shock waves are significantly stronger than and overtake the initial cylindrical shock waves originating from the bow shock envelope (see Silber, et al. (2018) for discussion). Depending on the atmospheric density and composition, and typically within the distance corresponding to the blast wave radius, the two waves coalesce into a stronger, unified shock wave, which continues to expand radially (Bronshten 1983, Tsikulin 1970). Sufficiently strong shock waves are often accompanied by secondary phenomena, including luminosity, plasma formation, hyperthermal chemical reactions, and the introduction of ablation-derived materials into the upper atmosphere (Ceplecha, et al. 1998, Silber, et al. 2018). The coupling of these processes with the surrounding atmosphere introduces variability in the amplitude and frequency content of the resulting acoustic signal. While this variability can complicate signal interpretation, it also produces distinct acoustic signatures that may be used to distinguish between different hypersonic sources.

In contrast, SRC reentries, such as OSIRIS-REx, are designed to maintain structural integrity and shape along their atmospheric trajectory. A lower velocity regime and the use of engineered materials minimize ablation and prevent fragmentation during hypersonic descent. Under nominal conditions, SRC-generated acoustic signatures are not contaminated by shock waves from more complex, ablating sources with highly varying and often poorly known parameters, which allows for more comprehensive interpretation of information content from unknown acoustic signals originating in the Earth's atmosphere. As such, these events serve as an excellent platform for study of the fundamentals of acoustic signals generated by non-ablational shock waves. Such signals may serve as a control or a well-defined baseline, enabling direct comparison with the signals from more complex sources.

The shock waves generated by SRCs are predominantly ballistic (expanding radially and perpendicularly to the axis of propagation). The comprehensive study of these shock waves is fundamentally important as they arise from the ballistic shock envelope of aerodynamically predictable and steady hypersonic motion of a SRC through the atmosphere. In principle, cylindrical shock waves, originating from a non-ablating body with a bow shock envelope are simpler to interpret. They are often modeled as conical wavefronts, with propagation dominated by the Mach cone geometry. These shocks exhibit predictable characteristics, allowing for more straightforward modeling and interpretation of the resulting infrasound signals (ReVelle, et al. 2006, Silber, et al. 2015). This comparison underlines the stochastic and highly variable nature of meteoroid-generated shocks versus the controlled and repeatable dynamics of SRC reentries.

The dense network of infrasound sensors deployed across Nevada and Utah enabled detailed analysis of the acoustic wavefield generated by the OSIRIS-REx SRC reentry as a function of source altitude, propagation path, and distance from the trajectory. The observed signals exhibited clear trends correlating with source altitude, with signals originating from higher altitudes displaying lower amplitudes and longer periods (Figure 3). This behavior aligns with theoretical predictions rooted in the physics of shock wave propagation through a rarefied and stratified atmosphere before it transitions into an acoustic wave. At higher altitudes, where the atmospheric density and pressure are significantly lower, the energy carried by the shock wave is spread over a larger volume, resulting in greater attenuation and reduced peak amplitudes. Fundamentally, the amplitude of the shock wave decreases approximately as a function of the inverse of the distance from the source, modified by atmospheric absorption factors dependent on altitude (e.g., Pierce, et al. 1990, Plooster 1968, ReVelle 1976, Silber, et al. 2018, Zel'dovich, et al. 2002). Accounting for the effect of attenuation, we have empirically demonstrated that the amplitude is predominantly governed by the source region rather than the propagation path, at least within the ranges investigated.

The observed N-wave (Dumond, et al. 1946) signatures (Figure 3) in the recorded signals were consistent across all stations, characterized by a prominent positive half-cycle (compression phase) followed by an extended negative phase (rarefaction phase). This is typical of shock-generated acoustic waves propagating in stratified atmospheres (Plooster 1970, ReVelle 1976). The asymmetry in the wave arises because nonlinear steepening during shock generation is gradually offset by relaxation as the wave propagates. The negative phase, in particular, becomes more diffuse due to shock attenuation and frequency-dependent relaxation effects. At higher altitudes, where air density and pressure are lower, these effects are amplified by greater geometric spreading and enhanced vibrational relaxation of atmospheric molecules, resulting in further attenuation and broadening of the waveform (e.g., Bass, et al. 1972, Bass, et al. 1995, Plooster 1968, Zel'dovich, et al. 2002).

The longer periods observed for signals originating from higher altitudes can be attributed to the distribution of acoustic energy from the attenuating shock wave front. In rarefied regions of the atmosphere, the initial shock envelope is larger and shock wave transitions into a weak shock or acoustic wave at larger radial distances, affecting temporal spreading. The acoustic signal is further influenced by the attenuation of high-frequency components, which are preferentially absorbed in low-density conditions due to molecular relaxation processes (e.g., Plooster 1970, Sutherland, et al. 2004). Consequently, the signals detected at ground stations from higher altitudes are dominated by lower frequencies, leading to longer periods. Energy dissipation also plays an important role. As the shock wave propagates downward, the cumulative energy flux diminishes with altitude due to viscous dissipation and thermal conduction. This attenuation is quantified by the absorption coefficient, which increases with decreasing density at higher altitudes (Sutherland, et al. 2004). The relationship between amplitude, period, and altitude in the near source region is thus governed by the combined effects of rarefaction, stratification, and molecular relaxation, which collectively influence the observable signal characteristics. These findings are consistent with prior studies on meteoroid-generated infrasound, where altitude-dependent signal attenuation were observed (ReVelle 1976, e.g., Silber, et al. 2014).

The clustering of signal parameters near the two distinct trajectory regions (Figure 6)—the peak heating zone (higher altitude) and the dynamic pressure maximum (lower altitude)—illustrates the interplay of competing effects on shock wave propagation. The amplitude correction applied through infraGA assumes a reference of 0 dB at an arbitrary distance of 1 km from the source, irrespective of altitude (Blom, et al. 2017), simplifying the representation of wave propagation dynamics. This assumption does not fully capture the complexities of source conditions, particularly the coupling of shock waves with the atmosphere at different altitudes. At higher altitudes, the shock wave is stronger at the source due to greater velocities and higher Mach numbers associated with the peak heating region. However, in the rarefied atmosphere, this is offset by an exponentially lesser number density of molecules and atoms in the shock front. Consequently, the energy coupling between the shock wave and the surrounding medium is less efficient, and the shock weakens as it propagates outward (Zel'dovich, et al. 2002). Thus, even at the conceptual 1 km reference distance, the corrected amplitude is lower, reflecting the combined effects of weaker energy coupling and greater geometric spreading in the first kilometer of propagation. In contrast, at lower altitudes near the dynamic pressure maximum, the shock wave is initially weaker due to lower velocities and Mach numbers. However, the denser atmosphere increases energy coupling, allowing more of the energy to propagate as acoustic waves (ReVelle 1976). Our observations are consistent with theoretical expectations of shock wave decay in stratified atmospheres and validate the ability of ground-based infrasound networks to capture the nuanced effects of altitude on shock wave characteristics. The amplitude trends encode information about both the source conditions and propagation dynamics, demonstrating the utility of infrasound sensing for trajectory analysis. The apparent higher amplitudes at lower altitudes, despite the initially weaker shocks, are consistent with theoretical predictions and provide a robust framework for studying shock wave dynamics in both controlled and natural atmospheric entry events.

The raytracing results showed excellent agreement between the predicted and observed arrival times, affirming the robustness of the propagation modeling methodology. The observed ballistic arrivals, corresponding to the shortest travel paths, provided reliable estimates of source altitudes along the SRC trajectory. This consistency supports the fundamental assumption that hypersonic shocks propagate in a ballistic manner, traveling near-perpendicularly to the trajectory with minimal deviation due to atmospheric effects in the near field. The agreement between raytracing-derived altitudes and the clustering of signal amplitude and period further supports the capability of distributed infrasound networks to capture trajectory-dependent variations in shockwave propagation. As mentioned earlier, stations closer to the trajectory consistently detected signals originating from lower altitudes, which exhibited higher amplitudes and shorter periods due to increased atmospheric density and stronger coupling of the shock wave to the medium. Conversely, stations further away captured signals from higher altitudes, characterized by lower amplitudes and longer periods as a result of reduced air density and greater propagation distances. These altitude-dependent signal characteristics validate the reliability of ballistic shock assumptions and demonstrate the feasibility of using infrasound to perform accurate trajectory reconstructions. This will be the subject for a future paper.

Finally, as the frequency of space debris reentries is expected to increase in the coming decades, understanding the differences between natural and artificial objects becomes increasingly critical. Unlike SRCs, space debris are composed of different materials, some of which may fragment or partially ablate during reentry, resulting in hybrid shock wave characteristics that lie between the extremes of meteoroids and engineered capsules. Studies like this, leveraging the controlled conditions of SRC reentries, provide a framework for interpreting the dynamic behavior of space debris and assessing its potential risks. Moreover, they contribute to the refinement of observational techniques and propagation models, offering tools to monitor both natural and artificial high-energy atmospheric entries effectively.

## Conclusions

The OSIRIS-REx Sample Return Capsule (SRC) reentry presented an unparalleled opportunity to investigate infrasound propagation dynamics and their dependence on source altitude and trajectory parameters. Utilizing a network of 39 ground-based infrasound sensors across Nevada and Utah, this study constitutes the most extensive geophysical campaign ever conducted for a hypersonic reentry. Raytracing results, rigorously validated against observed travel times, confirmed ballistic arrivals across all stations, with source altitudes inferred to range from 44 km to 62 km. These findings demonstrate the reliability of raytracing methodologies in characterizing hypersonic reentries. The infrasound signals exhibited a clear dependence on source altitude: higher-altitude sources produced signals with lower amplitudes, longer periods, and diminished high-frequency content, reflecting the interplay between shock wave strength, and the shock source region. Regression analysis further reinforced the strong correlations between signal parameters and source altitude. Our findings indicate that, after accounting for attenuation, the observed amplitude is primarily dictated by the source characteristics, with minimal influence from the propagation path within the ranges studied.

The controlled conditions of the OSIRIS-REx SRC reentry enabled a robust experimental investigation of atmospheric entry phenomena, offering benchmark data for direct comparisons with natural meteoroid events. Unlike the stochastic nature of meteoroid entries, SRC reentries provide a well-characterized analog for validating theoretical models under repeatable conditions. The comparison between predictable ballistic shocks from SRCs and ablational shocks from meteoroids illuminates the broader utility of SRC reentries in refining shock wave propagation models. This study demonstrates the utility of SRC reentries in advancing planetary defense strategies, refining shock wave propagation models, and improving global monitoring systems for high-energy atmospheric phenomena, including space debris reentries. With the anticipated rise in orbital traffic, these findings are crucial for addressing the increasing challenges of managing reentries in Earth's atmosphere. Moreover, the results provide a foundation for understanding atmospheric entry dynamics in planetary atmospheres, contributing to the study of high-energy phenomena in extraterrestrial environments.

SAND2025-03153O

## Data and Resources

InfraGA is freely available at: https://github.com/LANL-Seismoacoustics/infraGA. G2S is available at request through the National Center for Physical Acoustics (NCPA) at the University of Mississippi. Supplemental materials contain all tables generated as part of this work and are available online through the journal. The regression analysis was done using the proprietary graphing and analysis software package OriginPro®, developed by OriginLab®: https://www.originlab.com. The infrasound data were collected by an experimental network of the U.S. Government and are not currently available for release to the public.

## Declaration of Competing Interests

The authors declare no competing interests.

## Acknowledgements

The authors thank the two anonymous reviewers and the Associate Editor for their helpful feedback. This work was supported by the Nuclear Arms Control Technology (NACT) program at the Defense Threat Reduction Agency (DTRA). The authors thank M. Moreau (NASA) and S. Francis (Lockheed Martin Space) for sharing the EDL trajectory. The authors gratefully acknowledge the Board of Eureka County Commissioners, the Eureka Municipal Airport personnel, the Bureau of Land Management Ely District Office, and the Bureau of Land Management Salt Lake Field Office. This article has been authored by an employee of National Technology & Engineering Solutions of Sandia, LLC under Contract No. DE-NA0003525 with the U.S. Department of Energy (DOE). The employee owns all right, title and interest in and to the article and is solely responsible for its contents. The United States Government retains and the publisher, by accepting the article for publication, acknowledges that the United States Government retains a non-exclusive, paid-up, irrevocable, world-wide license to publish or reproduce the published form of this article or allow others to do so, for United States Government purposes. The DOE will provide public access to these results of federally sponsored research in accordance with the DOE Public Access Plan https://www.energy.gov/downloads/doe-public-access-plan. This paper describes objective technical results and analysis. Any subjective views or opinions that might be expressed in the paper do not necessarily represent the views of the U.S. Department of Energy or the United States Government.

SAND2025-03153O

## References

Ajluni, T., D. Everett, T. Linn, R. Mink, W. Willcockson, and J. Wood (2015). OSIRIS-REx, returning the asteroid sample, in *2015 IEEE aerospace conference*, IEEE, 1-15.

Anderson, J. D. (2000). Hypersonic and high temperature gas dynamics, American Institute of Aeronautics and Astronautics (AIAA).

Anderson, J. F., J. B. Johnson, D. C. Bowman, and T. J. Ronan (2018). The Gem infrasound logger and custom-built instrumentation, Seismological Research Letters **89** 153-164.

Bass, H. E., H.-J. Bauer, and L. B. Evans (1972). Atmospheric Absorption of Sound: Analytical Expressions, The Journal of the Acoustical Society of America **52** 821-825.

Bass, H. E., L. C. Sutherland, A. J. Zuckwerwar, D. T. Blackstock, and D. M. Hester (1995). Atmospheric absorption of sound: Further developments, The Journal of the Acoustical Society of America **97** 680-683.

Blom, P. (2014). GeoAc: Numerical tools to model acoustic propagation in the geometric limit, in *Software. Los Alamos National Laboratory*, Los Alamos National Laboratory, Seismoacoustic software.

Blom, P., and R. Waxler (2017). Modeling and observations of an elevated, moving infrasonic source: Eigenray methods, The Journal of the Acoustical Society of America **141** 2681-2692.

Bowman, D. C., E. A. Silber, M. R. Giannone, S. A. Albert, T. Edwards, F. K. D. Dugick, and R. S. Longenbaugh (2025). Acoustic Waves from the 20 April 2023 SpaceX Starship Rocket Explosion Traveling in the Elevated 'AtmoSOFAR' Channel, Geophysical Journal International ggaf091.

Bronshten, V. A. (1983). Physics of meteoric phenomena, Fizika meteornykh iavlenii, , Moscow, Izdatel'stvo Nauka, 1981 Dordrecht, D. Reidel Publishing Co, Dordrecht, Holland.

Brown, P., P. J. A. McCausland, M. Fries, E. Silber, W. N. Edwards, D. K. Wong, R. J. Weryk, J. Fries, and Z. Krzeminski (2011). The fall of the Grimsby meteorite—I: Fireball dynamics and orbit from radar, video, and infrasound records, Meteoritics & Planetary Science **46** 339-363.

Carr, C. G., C. Donahue, L. Viens, L. Beardslee, E. A. McGhee, and L. Danielson (2024). Detection of a space capsule entering Earth's atmosphere with distributed acoustic sensing (DAS), Seismological Research Letters.

Ceplecha, Z., J. Borovička, W. G. Elford, D. O. ReVelle, R. L. Hawkes, V. Porubčan, and M. Šimek (1998). Meteor Phenomena and Bodies, Space Science Reviews **84** 327-471.

SAND2025-03153O
Devillepoix, H. A. R., M. Cupák, P. A. Bland, E. K. Sansom, M. C. Towner, R. M. Howie, B. A. D. Hartig, T. Jansen-Sturgeon, P. M. Shober, S. L. Anderson, G. K. Benedix, D. Busan, R. Sayers, P. Jenniskens, J. Albers, C. D. K. Herd, P. J. A. Hill, P. G. Brown, Z. Krzeminski, G. R. Osinski, H. C. Aoudjehane, Z. Benkhaldoun, A. Jabiri, M. Guennoun, A. Barka, H. Darhmaoui, L. Daly, G. S. Collins, S. McMullan, M. D. Suttle, T. Ireland, G. Bonning, L. Baeza, T. Y. Alrefay, J. Horner, T. D. Swindle, C. W. Hergenrother, M. D. Fries, A. Tomkins, A. Langendam, T. Rushmer, C. O'Neill, D. Janches, J. L. Hormaechea, C. Shaw, J. S. Young, M. Alexander, A. D. Mardon, and J. R. Tate (2020). A Global Fireball Observatory, Planetary and Space Science **191** 105036.

Drob, D. P., J. M. Picone, and M. Garces (2003). Global morphology of infrasound propagation, Journal of Geophysical Research **108** 1-12.

Dumond, J. W. M., R. Cohen, W. K. H. Panofsky, and E. Deeds (1946). A Determination of the Wave Forms and Laws of Propagation and Dissipation of Ballistic Shock Waves, J. Acoust. Soc. America **18** 97-118.

Evans, L. B., H. E. Bass, and L. C. Sutherland (1972). Atmospheric Absorption of Sound: Theoretical Predictions, The Journal of the Acoustical Society of America **51** 1565-1575.

Fernando, B., C. Charalambous, C. Saliby, E. Sansom, C. Larmat, D. Buttsworth, D. Hicks, R. Johnson, K. Lewis, and M. McCleary (2024a). Seismoacoustic measurements of the OSIRIS-REx re-entry with an off-grid Raspberry PiShake, Seismica **3**.

Fernando, B., C. Charalambous, N. Schmerr, T. J. Craig, J. Wolf, K. Lewis, E. Sansom, C. Saliby, M. McCleary, and J. Inman (2024b). Array-based seismic measurements of OSIRIS-REx's re-entry.

Francis, S. R., M. A. Johnson, E. Queen, and R. A. Williams (2024). Entry, Descent, and Landing Analysis for the OSIRIS-REx Sample Return Capsule, in *46th Annual AAS Guidance, Navigation and Control (GN&C) Conference*, Breckenridge, CO.

Hetzer, C. H. (2024). The NCPAG2S command line client.

Ishihara, Y., Y. Hiramatsu, M.-y. Yamamoto, M. Furumoto, and K. Fujita (2012). Infrasound/seismic observation of the Hayabusa reentry: Observations and preliminary results, Earth, Planets and Space **64** 655-660.

Jenniskens, P., M. Gabadirwe, Q.-Z. Yin, A. Proyer, O. Moses, T. Kohout, F. Franchi, R. L. Gibson, R. Kowalski, E. J. Christensen, A. R. Gibbs, A. Heinze, L. Denneau, D. Farnocchia, P. W. Chodas, W. Gray, M. Micheli, N. Moskovitz, C. A. Onken, C. Wolf, H. A. R. Devillepoix, Q. Ye, D. K. Robertson, P. Brown, E. Lyytinen, J. Moilanen, J. Albers, T. Cooper, J. Assink, L. Evers, P. Lahtinen, L. Seitshiro, M. Laubenstein, N. Wantlo, P. Moleje, J. Maritinkole, H. Suhonen, M. E. Zolensky, L. Ashwal, T. Hiroi, D. W. Sears, A. Sehlke, A. Maturilli, M. E. Sanborn, M. H. Huyskens, S. Dey, K. Ziegler, H. Busemann, M. E. I. Riebe, M. M. M. Meier, K.

C. Welten, M. W. Caffee, Q. Zhou, Q.-L. Li, X.-H. Li, Y. Liu, G.-Q. Tang, H. L. McLain, J. P. Dworkin, D. P. Glavin, P. Schmitt-Kopplin, H. Sabbah, C. Joblin, M. Granvik, B. Mosarwa, and K. Botepe (2021). The impact and recovery of asteroid 2018 LA, Meteoritics & Planetary Science **56** 844-893.

Lauretta, D., S. Balram-Knutson, E. Beshore, W. Boynton, C. Drouet d'Aubigny, D. DellaGiustina, H. Enos, D. Golish, C. Hergenrother, and E. Howell (2017). OSIRIS-REx: sample return from asteroid (101955) Bennu, Space Science Reviews **212** 925-984.

Nemtchinov, I. V., V. V. Svetsov, I. B. Kosarev, A. P. Golub, O. P. Popova, V. V. Shuvalov, R. E. Spalding, C. Jacobs, and E. Tagliaferri (1997). Assessment of Kinetic Energy of Meteoroids Detected by Satellite-Based Light Sensors, Icarus **130** 259-274.

Pierce, A. D., and R. T. Beyer (1990). Acoustics: An Introduction to Its Physical Principles and Applications. 1989 Edition, The Journal of the Acoustical Society of America **87** 1826-1827.

Pilger, C., P. Hupe, P. Gaebler, and L. Ceranna (2021). 1001 Rocket Launches for Space Missions and Their Infrasonic Signature, Geophysical Research Letters **48** e2020GL092262.

Plooster, M. N. (1968). Shock Waves from Line Sources in *Shock*, 80.

Plooster, M. N. (1970). Shock Waves from Line Sources. Numerical Solutions and Experimental Measurements, Physics of Fluids **13** 2665-2675.

ReVelle, D. O. (1976). On meteor-generated infrasound, Journal of Geophysical Research **81** 1217-1230.

ReVelle, D. O., W. Edwards, and T. D. Sandoval (2005). Genesis—An artificial, low velocity "meteor" fall and recovery: September 8, 2004, Meteoritics & Planetary Science **40** 895-916.

ReVelle, D. O., and W. N. Edwards (2006). Stardust—An artificial, low-velocity "meteor" fall and recovery: 15 January 2006, Meteoritics & Planetary Science **42** 271-299.

Romig, M. F. (1964). The Physics of Meteor Entry, in *The RAND Corporation (declassified)*, DTIC Document, Santa Monica, California.

Sansom, E. K., H. A. Devillepoix, M.-y. Yamamoto, S. Abe, S. Nozawa, M. C. Towner, M. Cupák, Y. Hiramatsu, T. Kawamura, and K. Fujita (2022). The scientific observation campaign of the Hayabusa-2 capsule re-entry, Publications of the Astronomical Society of Japan **74** 50-63.

Silber, E., M. Ronac Giannone, D. C. Bowman, and S. Albert (2024a). Leveraging multi-station infrasound detections for characterization of high-altitude fireballs in *Infrasound Technology Workshop 2024 (ITW2024)*, CTBTO, Vienna, Austria.

Silber, E. A. (2024a). Perspectives and Challenges in Bolide Infrasound Processing and Interpretation: A Focused Review with Case Studies, Remote Sensing **16** 3628.

Silber, E. A. (2024b). The utility of infrasound in global monitoring of extraterrestrial impacts: A case study of the 23 July 2008 Tajikistan bolide, The Astronomical Journal **168**.

Silber, E. A., M. Boslough, W. K. Hocking, M. Gritsevich, and R. W. Whitaker (2018). Physics of meteor generated shock waves in the Earth's atmosphere – A review, Advances in Space Research **62** 489-532.

Silber, E. A., D. C. Bowman, and S. Albert (2023). A Review of Infrasound and Seismic Observations of Sample Return Capsules since the End of the Apollo Era in Anticipation of the OSIRIS-REx Arrival, Atmosphere **14** 1473.

Silber, E. A., D. C. Bowman, C. G. Carr, D. P. Eisenberg, B. R. Elbing, B. Fernando, M. A. Garces, R. Haaser, S. Krishnamoorthy, C. A. Langston, Y. Nishikawa, J. Webster, J. F. Anderson, S. Arrowsmith, S. Bazargan, L. Beardslee, B. Beck, J. W. Bishop, P. Blom, G. Bracht, D. L. Chichester, A. Christe, J. Clarke, K. Cummins, J. Cutts, L. Danielson, C. Donahue, K. Eack, M. Fleigle, D. Fox, A. Goel, D. Green, Y. Hasumi, C. Hayward, D. Hicks, J. Hix, S. Horton, E. Hough, D. P. Huber, M. A. Hunt, J. Inman, S. M. Ariful Islam, J. Izraelevitz, J. D. Jacob, J. Johnson, R. J. KC, A. Komjathy, E. Lam, J. LaPierre, K. Lewis, R. D. Lewis, P. Liu, L. Martire, M. McCleary, E. A. McGhee, I. N. A. Mitra, L. Ocampo Giraldo, K. Pearson, M. Plaisir, S. K. Popenhagen, H. Rassoul, M. Ronac Giannone, M. Samnani, N. Schmerr, K. Spillman, G. Srinivas, S. K. Takazawa, A. Tempert, R. Turley, C. Van Beek, L. Viens, O. A. Walsh, N. Weinstein, R. White, B. Williams, T. C. Wilson, S. Wyckoff, M.-Y. Yamamoto, Z. Yap, T. Yoshiyama, and C. Zeiler (2024b). Geophysical Observations of the 24 September 2023 OSIRIS-REx Sample Return Capsule Re-Entry, The Planetary Science Journal **5**.

Silber, E. A., and P. G. Brown (2014). Optical observations of meteors generating infrasound—I: Acoustic signal identification and phenomenology, Journal of Atmospheric and Solar-Terrestrial Physics **119** 116-128.

Silber, E. A., P. G. Brown, and Z. Krzeminski (2015). Optical observations of meteors generating infrasound: Weak shock theory and validation, Journal of Geophysical Research: Planets **120** 413-428.

Sutherland, L. C., and H. Bass, E. (2004). Atmospheric absorption in the atmosphere up to 160 km, J. Acoustic. Soc. Am. **115** 1012-1032.

Tsikulin, M. (1970). Shock waves during the movement of large meteorites in the atmosphere, DTIC Document AD 715-537, Nat. Tech. Inform. Serv., Springfield, Va

Zel'dovich, Y. B., and Y. P. Raizer (2002). Physics of shock waves and high-temperature hydrodynamic phenomena, Dover Publications.

**Supplemental materials**

**for**

**Along-trajectory acoustic signal variations observed during the hypersonic reentry of the OSIRIS-REx Sample Return Capsule**

Elizabeth A. Silber[1,*], Daniel C. Bowman[1,2,**]

[1]Sandia National Laboratories, Albuquerque, NM, 87144, USA; [2]Pacific Northwest National Laboratory, Richland, WA, 99354, USA

**Contents:**

Figures S1 – S4

Tables S1 – S4

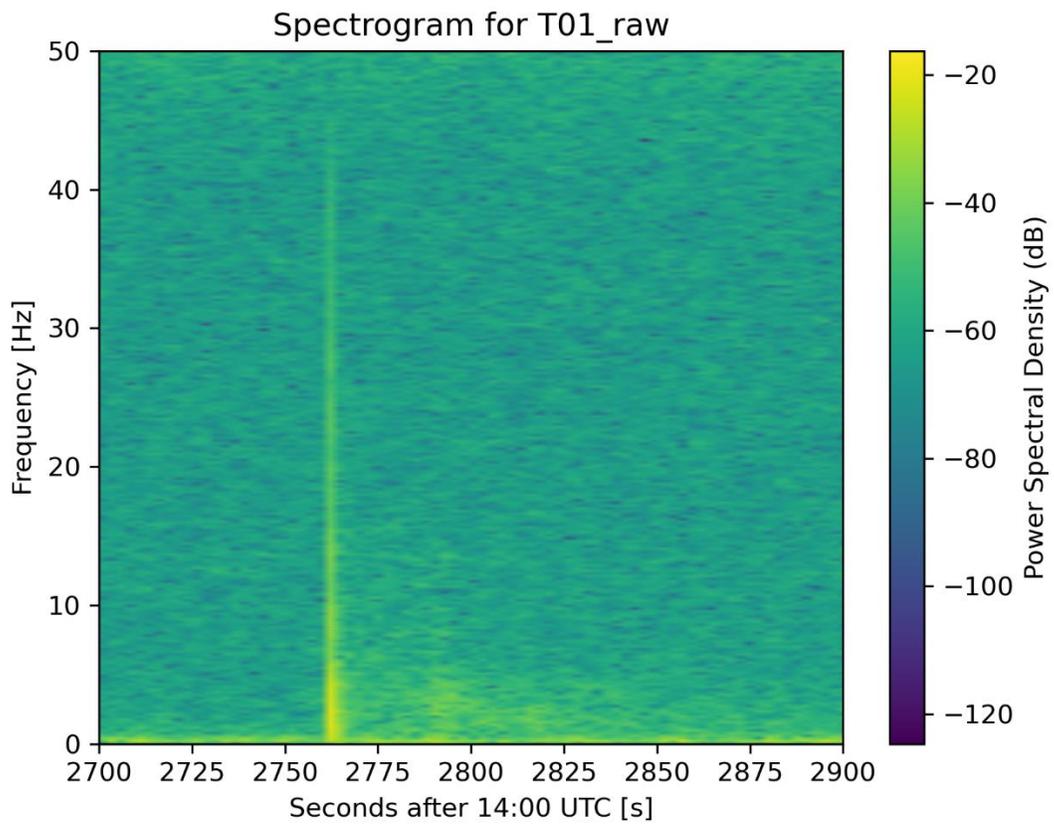

**Figure S1:** Spectrogram for station T1. The signal from the OSIRIS SRC can be seen as a 'pulse' at the ~2762 second mark.

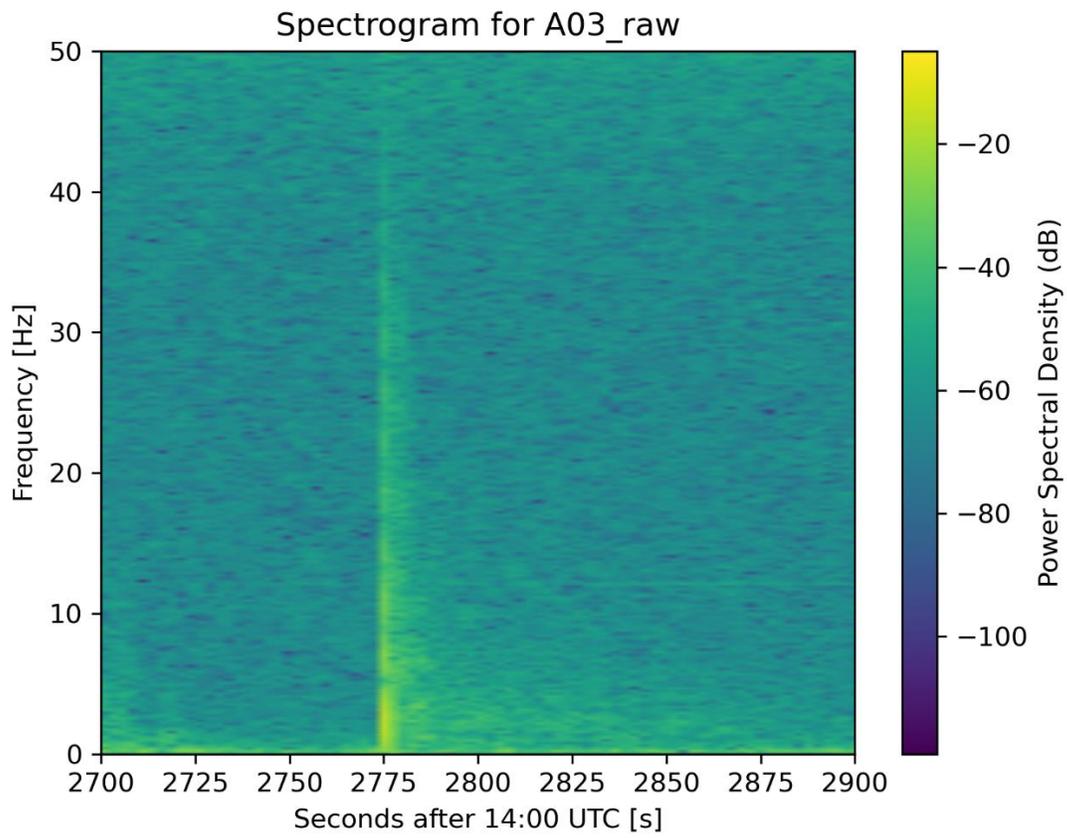

**Figure S2:** Spectrogram for station A3. The signal from the OSIRIS SRC can be seen as a 'pulse' at the ~2775 second mark.

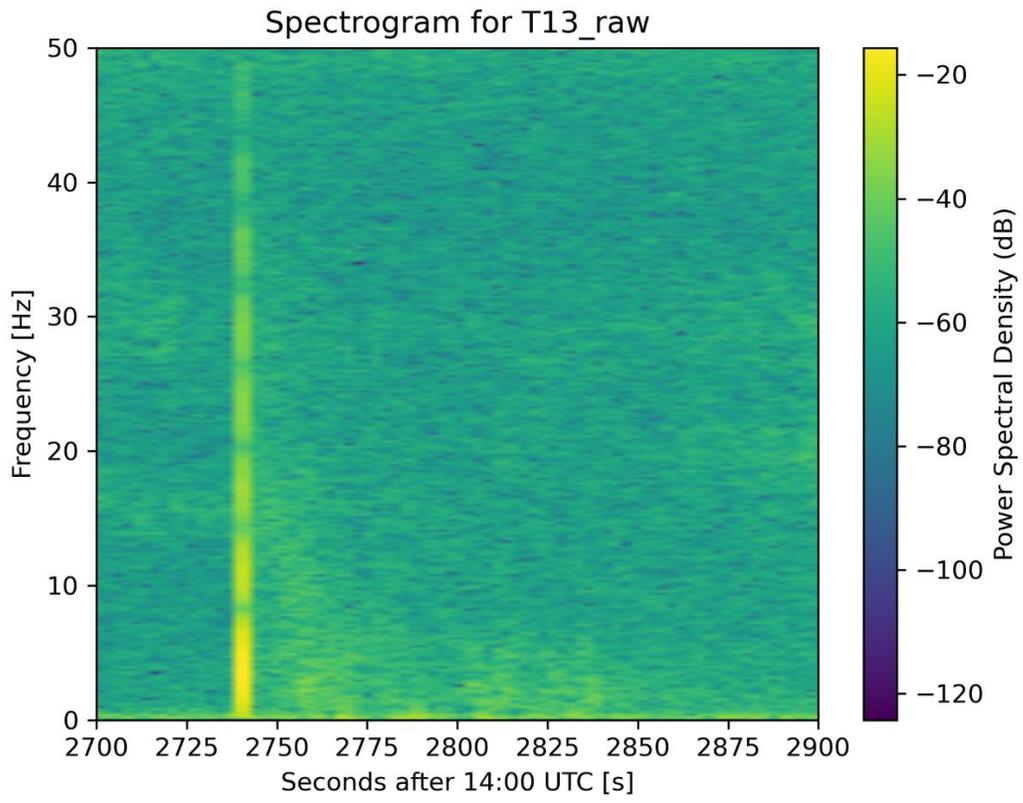

**Figure S3:** Spectrogram for station T13. The signal from the OSIRIS SRC can be seen as a 'pulse' at the ~2740 second mark.

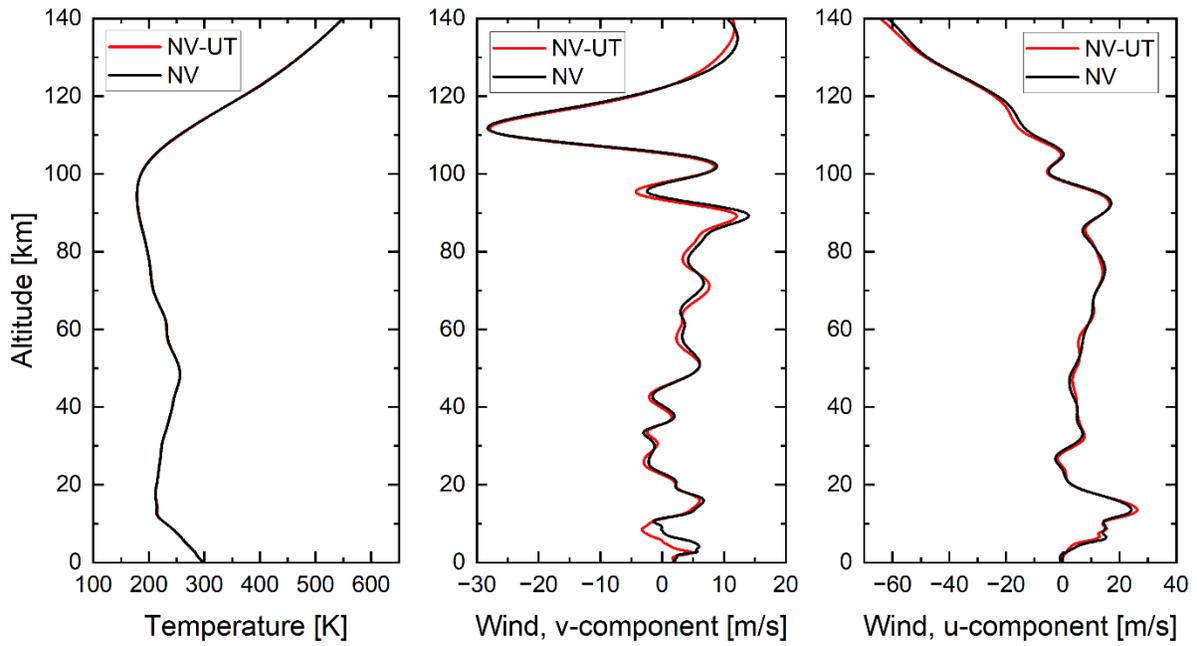

**Figure S4:** Atmospheric specifications (temperature and wind velocity) derived from G2S for two locations, Nevada (39.7480°N, -115.7730°E) (black line) and Utah-Nevada (40.1052°N, -113.9700°E) (red line). The temperature profiles from the two sites are nearly identical, hence the overlap.

**Table S1:** List of station particulars.

| Station ID | Station latitude [deg] | Station longitude [deg] | Station elevation [m] | Closest distance to trajectory [km] | Distance along track from origin [km] |
|---|---|---|---|---|---|
| A1 | 39.3766 | -115.8206 | 1811.66 | 35.32 | 39.66 |
| A2 | 39.4179 | -115.8125 | 1822.76 | 31.22 | 41.69 |
| A3 | 39.4574 | -115.8003 | 1819.37 | 27.44 | 44.20 |
| A4 | 39.5036 | -115.7845 | 1806.17 | 23.06 | 47.21 |
| A5 | 39.5479 | -115.7704 | 1823.36 | 18.82 | 49.69 |
| A7 | 39.6401 | -115.7789 | 1803.90 | 8.93 | 52.65 |
| A8 | 39.6810 | -115.7779 | 1789.83 | 4.68 | 54.11 |
| A9 | 39.7272 | -115.7679 | 1782.02 | 0.14 | 56.55 |
| A10 | 39.7488 | -115.7733 | 1800.86 | 2.29 | 57.03 |
| A11 | 39.7755 | -115.7652 | 1802.71 | 4.86 | 58.49 |
| A12 | 39.8155 | -115.7511 | 1822.06 | 8.64 | 61.37 |
| A13 | 39.8635 | -115.7320 | 1835.08 | 13.12 | 64.69 |
| A14 | 39.9170 | -115.7385 | 1904.63 | 18.90 | 66.11 |
| A15 | 39.9647 | -115.7465 | 1841.38 | 24.13 | 67.06 |
| A17 | 40.0523 | -115.7785 | 1807.32 | 34.21 | 67.99 |
| A18 | 40.0989 | -115.7832 | 1767.66 | 39.22 | 69.39 |
| A19 | 40.1440 | -115.7541 | 1729.97 | 43.12 | 72.64 |
| A20 | 40.1909 | -115.7434 | 1715.78 | 47.72 | 75.39 |
| A21 | 40.2318 | -115.7069 | 1769.82 | 50.97 | 79.92 |
| A22 | 40.2694 | -115.6817 | 1788.47 | 54.19 | 83.48 |
| T1 | 39.5414 | -116.3802 | 1868.12 | 1.86 | 0.00 |
| T2 | 39.5503 | -116.3368 | 1856.71 | 2.20 | 3.83 |
| T4 | 39.6023 | -116.1448 | 1835.73 | 2.34 | 21.53 |
| T7 | 39.7132 | -115.7629 | 1781.81 | 1.73 | 56.55 |
| T8 | 39.7437 | -115.6735 | 1828.43 | 1.09 | 64.69 |
| T10 | 40.0476 | -114.6406 | 1821.23 | 2.01 | 159.36 |
| T11 | 40.0636 | -114.5855 | 1898.51 | 2.18 | 164.36 |
| T13 | 40.0993 | -114.5496 | 1859.82 | 4.97 | 168.37 |
| T14 | 40.1821 | -114.0128 | 1602.72 | 0.71 | 214.87 |
| C1 | 39.9284 | -113.9995 | 1760.33 | 27.83 | 207.44 |
| C2 | 39.9713 | -113.9775 | 1714.80 | 23.89 | 210.39 |
| C3 | 40.0161 | -113.9731 | 1635.32 | 19.27 | 212.28 |
| C4 | 40.0611 | -113.9766 | 1642.66 | 14.43 | 213.36 |
| C5 | 40.1052 | -113.9701 | 1582.69 | 9.95 | 215.61 |
| C6 | 40.1518 | -113.9839 | 1579.06 | 4.67 | 216.20 |
| C7 | 40.1738 | -113.9970 | 1557.79 | 2.01 | 215.91 |
| C8 | 40.1945 | -113.9870 | 1543.37 | 0.08 | 217.36 |
| C9 | 40.2174 | -113.9901 | 1529.41 | 2.42 | 217.92 |
| C11 | 40.2881 | -113.9884 | 1527.69 | 9.83 | 220.69 |

**Table S2:** The summary of signal measurements.

| Station | Maximum amplitude [Pa] | Error in maximum amplitude [Pa] | Peak-to-peak amplitude [Pa] | Error in peak-to-peak amplitude [Pa] | Period [s] | Error in period [s] |
|---|---|---|---|---|---|---|
| A01 | 1.817 | 0.097 | 2.770 | 0.117 | 0.294 | 0.015 |
| A02 | 1.518 | 0.156 | 2.423 | 0.132 | 0.357 | 0.018 |
| A03 | 1.781 | 0.087 | 3.167 | 0.084 | 0.339 | 0.009 |
| A04 | 1.521 | 0.070 | 2.692 | 0.080 | 0.348 | 0.011 |
| A05 | 1.528 | 0.072 | 2.643 | 0.083 | 0.347 | 0.008 |
| A07 | 1.925 | 0.079 | 3.451 | 0.088 | 0.321 | 0.008 |
| A08 | 0.970 | 0.025 | 2.319 | 0.077 | 0.327 | 0.003 |
| A09 | 1.923 | 0.067 | 3.321 | 0.091 | 0.321 | 0.010 |
| A10 | 1.329 | 0.051 | 2.534 | 0.065 | 0.331 | 0.007 |
| A11 | 1.748 | 0.066 | 3.138 | 0.077 | 0.321 | 0.009 |
| A12 | 2.074 | 0.082 | 3.625 | 0.083 | 0.319 | 0.010 |
| A13 | 2.757 | 0.147 | 4.757 | 0.172 | 0.320 | 0.008 |
| A14 | 1.591 | 0.029 | 3.130 | 0.131 | 0.313 | 0.013 |
| A15 | 2.032 | 0.114 | 3.511 | 0.141 | 0.316 | 0.013 |
| A17 | 1.525 | 0.077 | 2.786 | 0.076 | 0.354 | 0.008 |
| A18 | 1.769 | 0.062 | 3.003 | 0.054 | 0.322 | 0.011 |
| A19 | 2.065 | 0.200 | 3.440 | 0.226 | 0.345 | 0.015 |
| A20 | 1.910 | 0.241 | 2.977 | 0.095 | 0.392 | 0.035 |
| A21 | 1.391 | 0.160 | 2.010 | 0.110 | 0.368 | 0.033 |
| A22 | 1.557 | 0.069 | 2.751 | 0.111 | 0.308 | 0.005 |
| T01 | 1.212 | 0.052 | 1.748 | 0.032 | 0.312 | 0.013 |
| T02 | 1.459 | 0.071 | 2.134 | 0.059 | 0.314 | 0.011 |
| T04 | 1.509 | 0.066 | 2.306 | 0.066 | 0.303 | 0.011 |
| T07 | 1.567 | 0.066 | 2.775 | 0.092 | 0.330 | 0.010 |
| T08 | 2.132 | 0.086 | 3.847 | 0.078 | 0.315 | 0.008 |
| T10 | 2.852 | 0.080 | 5.189 | 0.054 | 0.229 | 0.007 |
| T11 | 2.881 | 0.059 | 5.428 | 0.062 | 0.236 | 0.005 |
| T13 | 2.823 | 0.076 | 4.990 | 0.142 | 0.216 | 0.005 |
| T14 | 3.058 | 0.075 | 5.451 | 0.172 | 0.180 | 0.010 |
| C01 | 2.713 | 0.094 | 4.819 | 0.248 | 0.181 | 0.008 |
| C02 | 2.778 | 0.142 | 4.925 | 0.257 | 0.187 | 0.005 |
| C03 | 3.164 | 0.146 | 5.674 | 0.244 | 0.178 | 0.009 |
| C04 | 3.229 | 0.157 | 5.786 | 0.305 | 0.169 | 0.009 |
| C05 | 2.883 | 0.197 | 5.090 | 0.340 | 0.175 | 0.010 |
| C06 | 2.319 | 0.064 | 4.194 | 0.135 | 0.182 | 0.008 |
| C07 | 2.896 | 0.298 | 5.180 | 0.475 | 0.194 | 0.008 |
| C08 | 2.824 | 0.306 | 5.708 | 0.403 | 0.132 | 0.009 |
| C09 | 3.120 | 0.206 | 5.658 | 0.439 | 0.158 | 0.009 |
| C11 | 2.611 | 0.113 | 4.811 | 0.256 | 0.180 | 0.009 |

**Table S3:** The summary of raytracing results.

| Station | Source altitude [km] | Err source altitude [km] | Ground distance [km] | Err ground distance [km] | 3D distance [km] | Err 3D distance [km] |
|---|---|---|---|---|---|---|
| A01 | 58.91 | 0.30 | 35.69 | 0.39 | 67.34 | 0.11 |
| A02 | 58.76 | 0.30 | 31.61 | 0.41 | 65.12 | 0.13 |
| A03 | 58.54 | 0.29 | 27.89 | 0.46 | 63.21 | 0.12 |
| A04 | 58.29 | 0.29 | 23.58 | 0.56 | 61.21 | 0.11 |
| A05 | 58.09 | 0.29 | 19.41 | 0.61 | 59.52 | 0.13 |
| A07 | 57.86 | 0.27 | 9.97 | 1.10 | 56.95 | 0.12 |
| A08 | 57.71 | 0.27 | 6.42 | 1.73 | 56.32 | 0.12 |
| A09 | 57.44 | 0.26 | 4.70 | 2.40 | 55.91 | 0.12 |
| A10 | 57.49 | 0.26 | 4.58 | 2.11 | 55.91 | 0.11 |
| A11 | 57.36 | 0.27 | 6.46 | 1.62 | 55.96 | 0.13 |
| A12 | 57.14 | 0.26 | 9.71 | 1.11 | 56.17 | 0.11 |
| A13 | 56.86 | 0.27 | 13.94 | 0.87 | 56.77 | 0.11 |
| A14 | 56.76 | 0.27 | 19.45 | 0.59 | 58.21 | 0.12 |
| A15 | 56.69 | 0.29 | 24.57 | 0.47 | 60.10 | 0.13 |
| A17 | 56.64 | 0.29 | 34.53 | 0.33 | 64.80 | 0.13 |
| A18 | 56.54 | 0.29 | 39.51 | 0.29 | 67.53 | 0.12 |
| A19 | 56.21 | 0.30 | 43.41 | 0.28 | 69.66 | 0.12 |
| A20 | 56.01 | 0.30 | 47.98 | 0.25 | 72.46 | 0.13 |
| A21 | 55.66 | 0.30 | 51.24 | 0.24 | 74.36 | 0.12 |
| A22 | 55.41 | 0.30 | 54.44 | 0.20 | 76.42 | 0.12 |
| T01 | 62.16 | 0.30 | 4.81 | 2.59 | 60.54 | 0.12 |
| T02 | 61.84 | 0.29 | 4.91 | 2.42 | 60.23 | 0.12 |
| T04 | 60.39 | 0.29 | 4.87 | 2.31 | 58.80 | 0.13 |
| T07 | 57.51 | 0.27 | 4.47 | 2.41 | 55.96 | 0.11 |
| T08 | 56.80 | 0.28 | 4.75 | 2.28 | 55.23 | 0.14 |
| T10 | 49.41 | 0.24 | 4.05 | 1.85 | 47.80 | 0.12 |
| T11 | 48.99 | 0.26 | 4.29 | 1.94 | 47.32 | 0.12 |
| T13 | 48.66 | 0.24 | 6.22 | 1.26 | 47.23 | 0.12 |
| T14 | 44.26 | 0.27 | 4.90 | 2.09 | 42.98 | 0.10 |
| C01 | 45.24 | 0.32 | 28.19 | 0.34 | 51.82 | 0.13 |
| C02 | 44.89 | 0.32 | 24.31 | 0.41 | 49.55 | 0.13 |
| C03 | 44.69 | 0.32 | 19.77 | 0.51 | 47.38 | 0.13 |
| C04 | 44.54 | 0.32 | 15.09 | 0.69 | 45.48 | 0.12 |
| C05 | 44.31 | 0.33 | 10.91 | 0.97 | 44.11 | 0.13 |
| C06 | 44.24 | 0.32 | 6.35 | 1.60 | 43.16 | 0.12 |
| C07 | 44.29 | 0.32 | 4.41 | 2.12 | 43.01 | 0.13 |
| C08 | 43.98 | 0.28 | 4.86 | 2.17 | 42.77 | 0.11 |
| C09 | 44.04 | 0.32 | 4.77 | 2.06 | 42.82 | 0.11 |
| C11 | 43.76 | 0.33 | 10.79 | 0.97 | 43.60 | 0.13 |

**Table S4:** Regression results shown in Figure 7 (main text).

|  | Figure 7a | Figure 7b | Figure 7c | Figure 7d | Figure 7e | Figure 7f |
|---|---|---|---|---|---|---|
| **Signal parameter** | Period | Amplitude | Period | Amplitude | Period | Amplitude |
| **Distance parameter** | Source altitude | Source altitude | Total 3D distance | Total 3D distance | Ground distance from trajectory | Ground distance from trajectory |
| **Intercept** | -0.303 ± 0.040 | 13.247 ± 0.789 | -0.095 ± 0.038 | 9.467 ± 1.196 | 0.261 ± 0.016 | 3.109 ± 0.261 |
| **Slope** | 0.011 ± 7.413E-4 | -0.178 ± 0.015 | 0.007 ± 6.675E-4 | 0.110 ± 0.020 | 6.479E-4 ± 7.957E-4 | -0.006 ± 0.013 |
| **Residual sum of square** | 0.030 | 11.912 | 0.056 | 3725.145 | 2948.848 | 6649.596 |
| **Pearson's r** | 0.924 | -0.894 | 0.854 | -0.665 | 0.133 | -0.071 |
| **R-Square (COD)** | 0.854 | 0.799 | 0.729 | 0.443 | 0.018 | 0.005 |
| **Adj. R-Square** | 0.850 | 0.793 | 0.722 | 0.428 | -0.009 | -0.022 |